\documentstyle[twoside,fleqn,espcrc2]{article}

\input epsf

\newcommand{\Tr}{\mathop{\rm Tr}}
\newcommand{\Dslash}{/ \!\!\!\! D}
 
\newcommand{\LL}{\left\langle}
\newcommand{\RR}{\right\rangle}

\newcommand{\BE}{\begin{equation}}
\newcommand{\EE}{\end{equation}}
\newcommand{\BEA}{\begin{eqnarray}}
\newcommand{\EEA}{\end{eqnarray}}

\newcommand{\pbp}{\bar\psi\psi}
\newcommand{\Plaq}{\Box}
\newcommand{\gbeta}{6/g^2}
\newcommand{\step}{\Delta t}

\title{Thermodynamics for two flavor QCD\thanks{Presented by T.~Blum and C.~DeTar}}
\author{ C.~Bernard\address{Department of Physics, Washington University, St.~Louis, MO 63130, USA},
T.~Blum\address{Physics Department, Brookhaven National Laboratory, Upton, NY 11973, USA},
C.E.~DeTar\address{Physics Department, University of Utah, Salt Lake City, UT 84112, USA},
Steven~Gottlieb\address{Department of Physics, Indiana University, Bloomington, IN 47405, USA},
U.M.~Heller\address{SCRI, Florida State University, Tallahassee, FL 32306-4052, USA},
J.E.~Hetrick\address{Department of Physics, University of Arizona, Tucson, AZ 85721, USA},
L.~K\"arkk\"ainen\address{Nokia Research Center,
P.O.\ Box 100, FIN-33721 Tampere, Finland},
C.~McNeile$\,\null^{\rm c}$,
K.~Rummukainen$\,\null^{\rm d}$,
R.L.~Sugar\address{Department of Physics, University of California, Santa Barbara, CA 93106, USA},
D.~Toussaint$\,\null^{\rm f}$, 
and M.~Wingate\address{Department of Physics, University of Colorado, Boulder, CO 80309, USA},
} %end \author
\begin{document}

\begin{abstract}
We conclude our analysis of the $N_t=6$ equation of state (EOS) for two 
flavor QCD, first described
at last year's conference. We have obtained new runs
at $am_q=0.025$ and improved runs at $am_q=0.0125$. The results are extrapolated
to $m_q=0$, and we extract the speed of sound as well. We also 
present evidence for a restoration of the SU(2)$\times$SU(2) chiral symmetry
just above the crossover, but not of the axial U(1) chiral symmetry.
\end{abstract}

\maketitle
\section{INTRODUCTION}

The equation of state (EOS) is essential for phenomenological models
of heavy-ion collision experiments that seek to detect the quark-gluon
plasma (QGP).  It is also of cosmological interest since the strongly
interacting matter present in the early universe might have appeared
in such a state.

Following an initial EOS calculation at $N_t=4$ by some of
us~\cite{NT4}, we reported preliminary results for $N_t=6$ in
Melbourne~\cite{MEL}.  Since then, we have completed runs at
$am_q=0.025$ and extended runs at $am_q=0.0125$ to smaller step
size. With these additional simulations, we are able to fit our data
to empirical expressions, based on standard sigma model assumptions
with an O(4) or mean field critical point at zero quark mass.  The
fits allow an extrapolation to $am_q=0$ and permit a smooth
interpolation of the data, so, for example, the speed of sound can be
determined. We have also extended the $am_q=0.0125$ data to higher
temperature, roughly 250 MeV, where a plateau in the energy density is
now evident.

We refer the reader to Refs.~\cite{NT4,MEL} for details of our method.
There we describe how the pressure and energy density are found from 
derivatives of the
partition function with respect to the gauge coupling and bare quark
mass and a nonperturbative beta function.

\section{SIMULATIONS}

The analysis requires both hot ($12^3\times6$) and cold ($12^4$)
lattices.
Each hot (cold) simulation is at least 1800 (800) time units long after 
equilibration. 
Near the crossover region the simulations were extended to more than 3000 units
for the hot lattices. 
The range of gauge couplings and masses 
corresponds roughly to physical
temperatures, $0<T<250$ MeV (based on the $\rho$ mass~\cite{NT4}), 
and mass ratios, $0.3<m_{\pi}/m_\rho<0.7$.
We have also performed simulations at various 
$R$ algorithm
step sizes $\step$ to extrapolate results to $\step=0$. 
For instance, extrapolating the step size from 0.005--0.01 to zero 
produces a cumulative change in the pressure of 5\% at $6/g^2 = 5.6$.

\section{RESULTS}

\begin{figure}
\vbox{\hskip -0.25in\epsfxsize=3.0in \epsfbox[0 0 4096 4096]{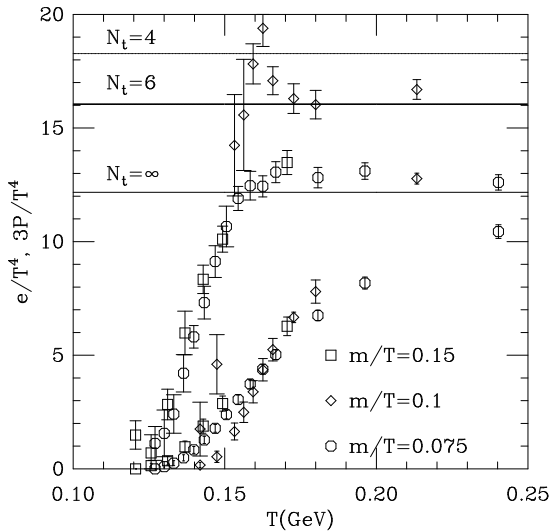} }
\vskip -0.5in
\caption{ \label{eos} The EOS. The diamonds indicate an earlier 
result on $N_t=4$ lattices.}
\end{figure}

In Fig.~\ref{eos} we compare the $N_t=6$ EOS with the $N_t=4$ result and 
the continuum and lattice Stefan-Boltzmann laws. Here we have converted to a physical
temperature scale. There is an apparent large finite
size effect which is expected from the free lattice theory. For $N_t=6$ we see
that the results depend weakly on the quark mass.
{}From the location of the maximum in the slope of $\LL\Plaq\RR$ or $\LL\pbp\RR$ with 
respect to $\gbeta$, we conclude that the pseudocritical temperature of the transition is 
roughly 140 MeV (for both $am_q=0.025$ and 0.0125). 
Figure~\ref{eos} shows that the energy density is already 
substantial at this point.
At high temperature the energy density has leveled off dramatically and its expected
approach to the free lattice result is slow.

\begin{figure}
\vbox{ \hskip-.25in\epsfxsize=3.0in \epsfbox[0 0 4096 4096]{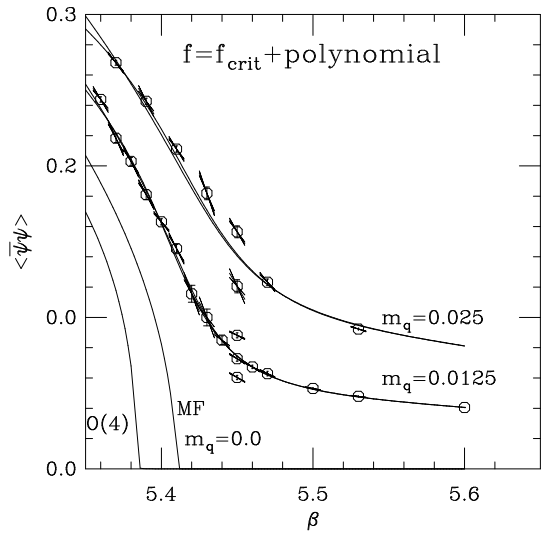}.  }
\vskip -0.5in
\caption{\label{order param} Fit to $\pbp$ and its derivative with respect to $\gbeta$.
The difference between O(4) and mean field is not discernible with the present data. Extrapolation
to $m_q=0$, however, gives a different critical coupling.}
\end{figure}

To obtain a smooth interpolation of the above results and extrapolate
to $m_q\approx 0$ (the physical value), we assume a second order
phase transiton at $m_q=0$ (but see Ukawa's review\cite{UKAWA} for cautionary
remarks).  The nonanalytic part of the free energy obeys the scaling
relation
\BEA
f_{\rm crit}(t,h) &=& b^{-d} f_{\rm crit}(b^{y_t}t,b^{y_h}h) 
\EEA
where $t$ and $h$, the usual scaling variables,
are proportional to $\gbeta-(\gbeta)_c$ and $am_q$, respectively.
Both the critical exponents, $y_h$ and $y_t$, and the scaling
function $f_{\rm crit}(t,h)$ are universal. Two flavor QCD has long been expected to be 
in the same universality class as the 3d O(4) Heisenberg magnet~\cite{O4}. 
While the O(4) critical exponents are known~\cite{O4EX}, the scaling function is not. Therefore
we have performed new simulations of the O(4) magnet to determine the scaling function.
Recently it has been argued that the transition may be mean field~\cite{KK},
for which the entire form of the scaling part of the free energy is known~\cite{PFEUTY}.

In Fig.~\ref{order param} we show preliminary fits of $\pbp$ and its derivative to the mean field 
and O(4) scaling functions plus a polynomial in $am_q$ and $6/g^2$. 
For the quark masses used in our simulations, the fits are indistinguishable. 
However, the respective extrapolations to $m_q=0$ are quite different. In Fig.~\ref{order param}
the mean field and O(4) critical couplings correspond to {\em zero mass} temperatures 
$T_c\approx 150$ and 160 MeV, respectively.
When $\LL\Plaq\RR$ and its derivative are included in the fits, these values are shifted
down by about 10 MeV.
The above indicates that present lattice simulations may still be too far from the
scaling region and smaller quark masses are required to see the true 
scaling behavior. 

\begin{figure}[hbt]
\vbox{\hskip -0.25in\epsfxsize=3.0in \epsfbox[0 0 4096 4096]{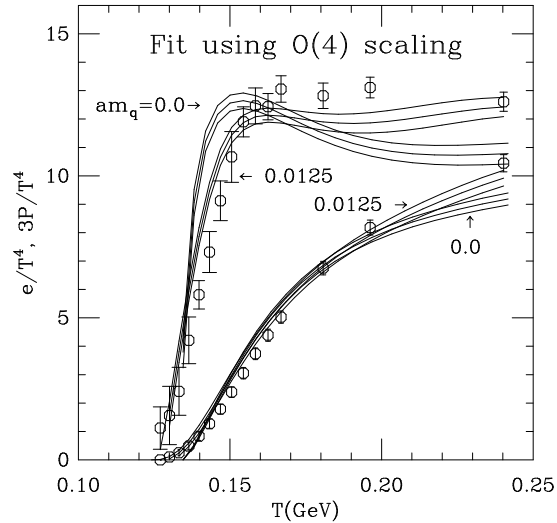} }
\vskip -0.5in
\caption{ \label{eosfit} The EOS extrapolated to $am_q=0.0$. The
fit and data at $am_q=0.0125$ are also shown.  The triplet of fit
curves at both masses, as labeled, indicates the central value and one
standard deviation spread resulting from the statistical uncertainty
in the fit.
\vskip -0.35in}
\end{figure}

\begin{figure}[hbt]
\vbox{\hskip -0.25in\epsfysize=1.5in\epsfxsize=3.0in \epsfbox[0 0 4096 2096]{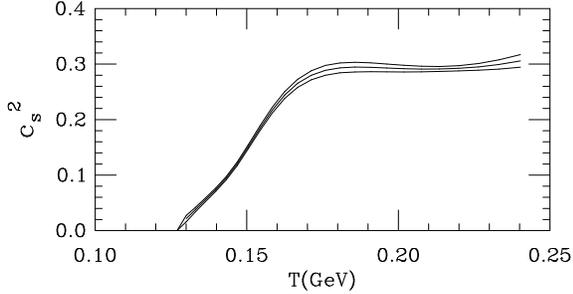} }
\vskip -0.5in
\caption{ \label{sound} The speed of sound for $am_q=0.0125$. The three curves
indicate the central value and the one standard deviation spread resulting
from the statistical uncertainty in the fit.}
\end{figure}

An extrapolation of the EOS to $m_q=0$ is shown in Fig.~\ref{eosfit}. It
is compared with the $am_q=0.0125$ result, which reproduces the data reasonably well.
Note that here we fit $\LL\Plaq\RR$, $\LL\pbp\RR$ and their 
derivatives with respect to
$\gbeta$ simultaneously. The appearence of the bump in the energy density just after 
the transition is probably
an artifact of the extrapolation (at $m_q=0$, the corresponding region of $\gbeta$ lies 
below the values of the coupling where we have done simulations). From Fig.~\ref{eosfit} we
again see a weak dependence on the quark mass.
In Fig.~\ref{sound} we show the speed of sound calculated
from this fit. The low temperature part of the curve is not trustworthy since the
derivatives of the energy density and pressure are poorly known in this region. In fact, we expect
the hadron gas below the transition to have a nonzero speed of sound,
which then dips down at the transition.

\section{CHIRAL SYMMETRY RESTORATION}

Sigma model analogies suggest that in the chiral limit of zero quark
mass in the quark-gluon plasma phase, the SU(2)$\times$SU(2) chiral
symmetry is exact, but the anomalous U(1)$_A$ axial symmetry may
remain broken\cite{O4}.  Early efforts to assess the status of these
symmetries looked for the expected degenerate multiplets in screening
masses.  However, because these studies did not include
difficult-to-measure quark-line disconnected graphs, they were
inconclusive\cite{SHURYAK}.  The large data sample generated in the
EOS study makes it feasible to reexamine this question.  Knowing the
answer gives important insight into the mechanics of the phase
transition.  The Columbia group has also pursued this question
\cite{CHANDRA-CHRIST}, the Bielefeld group has reported preliminary 
results \cite{BOYD}, as has Lagae \cite{LAGAE}

To address this question we focused on susceptibilities for the
following hadron channels: $f_0$ (also known as $\sigma$ or $\bar \psi
\psi$), $a_0$ ($\delta$), and $\pi$.  For example the $f_0$
susceptibility is constructed from the point-source correlator as
follows:
\BE
  \chi_{f_0} = \int d^4x \langle f_0(x) f_0(0) \rangle 
	- \langle f_0(0) \rangle^2
\EE
{}From these susceptibilities, we define the order parameters
\BE
   \chi_{U(1)}  = \chi_\pi   - \chi_{a_0}; \ \ \ \ 
   \chi_{SU(2)} = \chi_{f_0} - \chi_\pi.
\EE
If the order parameter is nonzero in the chiral limit, the
corresponding symmetry is broken.

The $f_0$ susceptibility can be decomposed into quark-line-connected
and disconnected parts as $\chi_{f_0} = \chi_{\rm conn} + \chi_{\rm disc}$.
The connected part is identified with $\chi_{a_0}$, so is obtained
from the point hadron correlator using traditional methods.  The
disconnected part is the variance from configuration to configuration
of the space-time volume average of $f_0 = \bar\psi\psi$.  The latter is
conventionally measured from $\Tr (\Dslash + m_q)^{-1}$, which we
estimated using multiple random sources on each configuration.
Finally, the pion susceptibility $\chi_{\pi}$ can also be obtained
from $\langle f_0 \rangle/m$.

For this study we fixed $6/g^2 = 5.45$ and varied the quark mass from
$am_q = 0.0075$ to $0.025$, corresponding to the high temperature
phase.  The crossover to the low temperature phase occurs at a
slightly higher mass, and the chiral limit corresponds to a
temperature of approximately $1.2 T_c(ma=0)$.

\begin{figure}
\vbox{\vskip-0.5in
\hskip -0.25in\epsfxsize=3.0in \epsfbox[0 0 4096 4096]{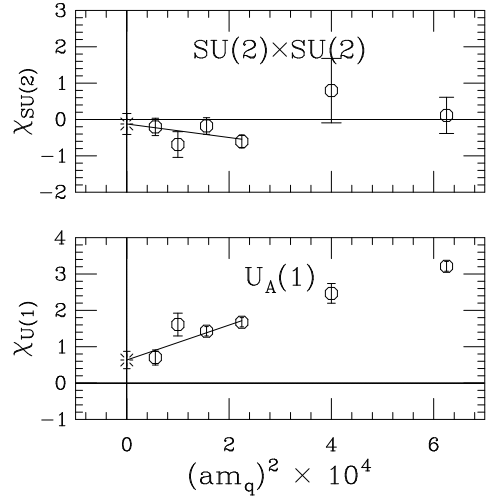} }
\vskip -0.5in
\caption{ \label{chiral} Chiral symmetry order parameters for $T > T_c$, extrapolated
to zero quark mass (burst symbol). The two high mass points closest to the
crossover were omitted from the fit.}
\end{figure}

Preliminary results are shown in Fig.~\ref{chiral}.  In the two-flavor
staggered fermion scheme the fermion determinant is rigorously even in
the quark mass.  In consequence the order parameters are also even
in $am_q$.  For $T > T_c$ there are no known infrared singularities,
even in the chiral limit.  Thus we have indicated a linear
extrapolation in $(am_q)^2$, leading to a result consistent with the
sigma model scenario: a restoration of $SU(2)\times SU(2)$ but not of
$U(1)_A$ (approximately $2\sigma$).

This work was supported by the US DOE and NSF.  Computations were done
at the San Diego Supercomputer Center, the Cornell Theory Center, Indiana
University, and the University of Utah.

\end{document}